\documentclass[prl,twocolumn,showpacs,floatfix,amsmath,nofootinbib,amssymb,floatfix]{revtex4}
\usepackage{graphicx,color,dcolumn,booktabs,bm}
\usepackage{longtable,lscape}
\usepackage{txfonts}
\usepackage{overpic}
\usepackage{amssymb}
\usepackage{indentfirst}
\usepackage{feynmf}   
\usepackage{slashed}  
\usepackage{cases}
\usepackage{color}
\usepackage{multirow}
\usepackage{epstopdf}
\usepackage{float}
\usepackage{graphicx,color,dcolumn,booktabs,bm}
\usepackage[colorlinks, citecolor=blue,anchorcolor=red,menucolor=red, linkcolor=red,filecolor=red,runcolor=red,urlcolor=blue,frenchlinks=red]{hyperref}

\graphicspath{{Figures/}} %

\begin{document}

\title{Direct Evidence for the $\bar{D}D^*/D\bar{D}^*$ Molecular Nature of $G(3900)$ Through Triangular Singularity Mechanisms}

\author{Yin Huang$^{1,2}$}
\author{Xurong Chen$^{2,3,4,5}$} \email{xchen@impcas.ac.cn}
\affiliation{$^{1}$School of Physical Science and Technology, Southwest Jiaotong University, Chengdu 610031,China}
\affiliation{$^{2}$Southern Center for Nuclear-Science Theory (SCNT), Institute of Modern Physics, Chinese Academy of Sciences, Huizhou 516000, China}
\affiliation{$^{3}$Institute of Modern Physics, Chinese Academy of Sciences, Lanzhou 730000, China }
\affiliation{$^{4}$School of Nuclear Science and Technology, University of Chinese Academy of Sciences, Beijing 100049, China}
\affiliation{$^{5}$State Key Laboratory of?Heavy Ion Science and Technology, Institute of Modern Physics, Chinese Academy of Sciences, Lanzhou 730000, China}

\begin{abstract}
The exotic hadron \( G(3900) \), initially observed in the process \( e^+ e^- \to D\bar{D} \), has been further supported by analyses from the BESIII Collaboration, which classify it as a \( P \)-wave molecular state of \( D^{+}D^{*-}/D^{-}D^{*+} \). However, theoretical discussions raise concerns about its status as a true particle, emphasizing the need for additional studies. In this Letter, we employ the triangular singularity mechanism to investigate \( G(3900) \) across various reaction channels, allowing us to produce significant peaks without relying on the existence of a real particle.  We identify \( X(4020) \), \( Y(4320) \), and the tentative \( X(4014) \) as potential sources of these peaks via decay processes to \( \gamma G(3900) \) or \( \pi G(3900) \). We stress the importance of experimental explorations of \( e^+ e^- \to X/Y \to \gamma(\pi) G(3900) \), which are essential for confirming the molecular composition of \( D^{+}D^{*-}/D^{-}D^{*+} \) and refining the mass measurement of \( G(3900) \).
\end{abstract}

\maketitle
\textit{Introduction}--The exploration of hadron structure and the quest for new hadronic states are fundamental endeavors in particle physics. To date, numerous hadrons have been experimentally established~\cite{ParticleDataGroup:2022pth}, many of which can be categorized as conventional quark states, including mesons formed from quark-antiquark pairs~\cite{Godfrey:1985xj} and baryons constructed from three quarks~\cite{Koniuk:1979vy, Isgur:1978wd}.  However, certain hadrons demonstrate more intricate internal architectures, prompting their classification as exotic states~\cite{Guo:2017jvc}. Among the diverse array of theoretical frameworks, the hadron-hadron molecule model serves as a prominent interpretation of these exotic states, as molecular structures are not merely abstract notions but are realized in nature-most notably illustrated by atomic nuclei, which consist of protons and neutrons. The $\Lambda(1405)$ is another well-accepted molecular candidate, widely recognized as being predominantly a $\bar{K}N$ bound state~\cite{Oset:1997it, Oller:2000fj, Nemoto:2003ft, Hall:2014uca}.
Lattice QCD studies~\cite{Nemoto:2003ft, Hall:2014uca} provide compelling evidence supporting the identification of $\Lambda(1405)$ as a molecular state involving $\bar{K}N$ and its coupled channels. A landmark discovery in this domain is the exotic hadron $X(3872)$~\cite{Belle:2003nnu}, frequently interpreted as a $\bar{D}D^{*}$ molecule~\cite{Brambilla:2019esw, Chen:2022asf, Meng:2022ozq}. Leveraging heavy quark symmetry, it is anticipated that additional molecular states containing $\bar{D}^{*}D^{*}$, $\bar{B}^{*}B$, and $\bar{B}^{*}B^{*}$ components should exist, and such hadrons have indeed been observed, including the $Z_c(4020/4025)$~\cite{BESIII:2013mhi, BESIII:2013ouc, BESIII:2014gnk}, $Z_b(10610)$, and $Z_b(10650)$~\cite{Belle:2011aa}. Noteworthy candidates for molecular states include the hidden-charm pentaquark states $P_c(4312)$, $P_c(4380)$, $P_c(4440)$, $P_c(4457)$, $P_{cs}(4338)$, and $P_{cs}(4459)$, identified by the LHCb Collaboration~\cite{LHCb:2015yax, LHCb:2016ztz, LHCb:2016lve, LHCb:2019kea, LHCb:2020jpq, LHCb:2022ogu}. These states can be effectively described as $\bar{D}^{(*)}\Sigma_c^{(*)}$ or $\bar{D}^{(*)}\Xi_c$ molecular configurations~\cite{Chen:2019bip, Guo:2019fdo, Xiao:2019aya, He:2019ify, Xiao:2019mvs, Roca:2015dva, Chen:2015moa, Chen:2015loa, Yang:2015bmv, Huang:2015uda, Du:2019pij}.

The discovery of such particle states has invigorated theoretical and experimental pursuits of additional hadrons containing heavy quarks. Recently, an exotic hadron, $G(3900)$, was identified near the mass of the $X(3872)$, which was first established by the Belle Collaboration in 2003~\cite{Belle:2003nnu, Brambilla:2019esw, Chen:2022asf, Meng:2022ozq}. The BESIII Collaboration has since observed this state in their analysis of the $e^{+}e^{-} \to D\bar{D}$ process~\cite{BESIII:2024ths}, reporting its mass and width as $\mathrm{M}_{G(3900)} = 3872.5 \pm 14.2 \pm 3.0 \ \mathrm{MeV}$ and $\mathrm{\Gamma}_{G(3900)} = 179.7 \pm 14.1 \pm 7.0 \ \mathrm{MeV}$. Given that the central mass of $G(3900)$ closely aligns with that of $X(3872)$, along with its inferred \( c\bar{c}q\bar{q} \) composition from the observed $D\bar{D}$ channel, it is plausible to consider the newly discovered $G(3900)$ as a high orbital angular momentum molecular state involving $\bar{D}D^{*}/D\bar{D}^{*}$. In particular, Ref.~\cite{Lin:2024qcq} designates it as the inaugural $P$-wave $\bar{D}D^{*}/D\bar{D}^{*}$ dimeson state, derived from solutions to the complex-scaled Schr\"{o}dinger equation in momentum space utilizing a $P$-wave $\bar{D}D^{*}/D\bar{D}^{*}$ one-boson exchange interaction potential.

The experimental signature of the candidate state $G(3900)$ was initially detected by the BABAR Collaboration in 2007~\cite{BaBar:2006qlj} and subsequently by the Belle Collaboration in 2008~\cite{Belle:2007qxm} during the analysis of the $e^{+}e^{-} \to D\bar{D}$ reaction. At that time, this enhancement observed near 3900 MeV was not perceived as a distinct hadronic state, as it could be attributed to the opening of the $D\bar{D}^* + D^* \bar{D}$ decay channel and the influence of a node in the $\psi(3S)$ wave function~\cite{Eichten:1979ms}. However, the investigations in Refs.~\cite{Uglov:2016orr, Nakamura:2023obk, Du:2016qcr} have highlighted the challenges in accurately modeling the line shape of the $e^{+}e^{-} \to \bar{D}D$ reaction without considering the $G(3900)$ state. After the $G(3900)$ state was reported again by BESIII~\cite{BESIII:2024ths}, the authors in Refs.~\cite{Salnikov:2024wah, Husken:2024hmi} combined previous and latest data for analysis, and once again concluded that the state reported by BESIII~\cite{BESIII:2024ths} does not represent a genuine hadronic molecule.
 The discrepancies between the results of Refs.~\cite{Salnikov:2024wah, Husken:2024hmi} and the BESIII observations~\cite{BESIII:2024ths}-alongside previous studies~\cite{Lin:2024qcq, Uglov:2016orr, Nakamura:2023obk, Du:2016qcr}-prompt critical inquiries into the nature of $G(3900)$:
\begin{itemize}
    \item Is $G(3900)$ a genuine hadronic state?
    \item If so, could it represent a $P$-wave $\bar{D}D^{*}/D\bar{D}^{*}$ molecular resonance, as suggested in Refs.~\cite{Lin:2024qcq, Du:2016qcr}?
\end{itemize}
To resolve these pivotal questions, it is essential to detect the $G(3900)$ in a reaction that is independent of the $e^{+}e^{-} \to \bar{D}D$ channel, with the objective of elucidating its molecular characteristics related to $\bar{D}D^{*}/D\bar{D}^{*}$.

In particle physics, the identification of states is typically achieved through the observation of resonance peaks in designated reaction pathways. However, given the incomplete comprehension of Quantum Chromodynamics (QCD), the mechanisms responsible for the enhancement of these resonance peaks remain ill-defined. Triangular singularities, conversely, can manifest as distinctive peaks in specific hadronic processes~\cite{Guo:2017jvc, Achasov:1989ma, Bityukov:1986yd, BESIII:2012aa, Wu:2011yx, Mikhasenko:2015oxp, Guo:2014iya, Guo:2015umn, Guo:2010ak, Guo:2019twa, Liu:2015taa, Sakthivasan:2024uwd, Guo:2019qcn}, thus offering promising experimental signals for the identification of new hadronic states. Moreover, triangular singularities can be instrumental in determining whether a given state is molecular in nature~\cite{Guo:2017jvc, Guo:2015umn, Liu:2015taa, Sakthivasan:2024uwd, Guo:2019qcn}. Consequently, in this Letter, we propose investigating the $G(3900)$ as a potential hadronic state in alternate reaction processes, such as $e^{+}e^{-}$ collisions, utilizing the triangular singularities mechanism. Upon confirmation, this approach may simultaneously assess its molecular resonance as a $\bar{D}D^{*}/D\bar{D}^{*}$ complex.

\begin{figure}[http]
\begin{center}
\includegraphics[bb=50 560 1050 710, clip, scale=0.42]{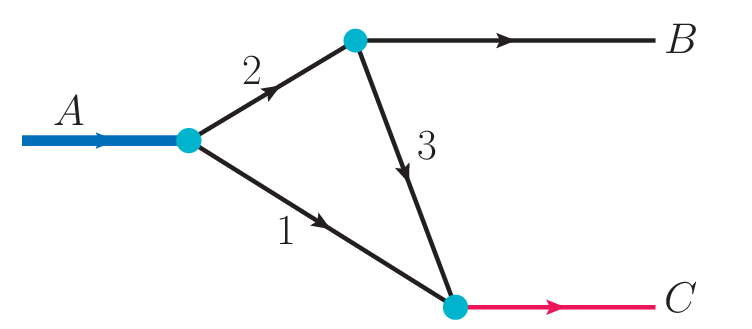}
\caption{The Feynman diagrams for the decay of \( A \) to \( BC \) via a triangle loop involve intermediate states 1, 2, and 3.
In this process, particle \( A \) decays into particles 1 and 2, particle 2 then decays into particles 3 and \( B \), and finally,
particles 1 and 3 interact to produce the final state \( C \).
}\label{cc1}
\end{center}
\end{figure}

\textit{Our Strategies}--Triangular singularity was first reported by Landau~\cite{Karplus:1958zz, Landau:1959fi} and discussed in detail in Refs.~\cite{Guo:2017jvc,Achasov:1989ma,Bityukov:1986yd,BESIII:2012aa,Wu:2011yx,Mikhasenko:2015oxp,Guo:2014iya,Guo:2015umn,Guo:2010ak,Guo:2019twa, Liu:2015taa,Sakthivasan:2024uwd,Guo:2019qcn}. Such singularities imply that peaks observed in the invariant mass distribution or scattering cross-section
of reactions do not necessarily correspond to genuine particle states. These peaks arise from a triangular decay process \( A \to BC \), involving three
intermediate particles (1, 2, and 3), as depicted in Fig.~\ref{cc1}, when the masses and energies of these particles satisfy the Landau
equation~\cite{Karplus:1958zz, Landau:1959fi}:
\begin{align}
1 + 2y_{12}y_{23}y_{13} = y_{12}^2 + y_{23}^2 + y_{13}^2.\label{eq1}
\end{align}
Where \( y_{ij} = (m_i^2 + m_j^2 - p_{ij}^2)/(2m_i m_j) \), with \( m_i \) \((i=1,2,3)\) representing the masses of intermediate particles, and
\( p_{ij}^2 = (p_i + p_j)^2 = m_{ij}^2 \) being the four-momentum of the \( ij \) pair, which corresponds to the external particle.  Coleman and
Norton's careful analysis of Eq.~\ref{eq1} revealed that the presence of triangular singularities, which
appears in Fig.~\ref{cc1}, depends on whether the decay process is classical and whether all three internal particles can be simultaneously on-shell
and collinear in the rest frame of the decaying particle $A$, a result known as the Coleman-Norton theorem~\cite{Coleman:1965xm}.  Specifically, the
triangular singularity can only occur when the initial particle \(A\) and the final
particle \(C\) satisfy the following conditions~\cite{Schmid:1967ojm}:
\begin{align}
    &M^2_A \in \left[(m_1 + m_2)^2, m_1^2 + m_2^2 + 2m_1 \frac{m_3^2 + m_2^2 - M_B^2}{2m_3}\right],\label{eq2}\\
    &M_C^2 \in \left[(m_1 + m_3)^2, m_1^2 + m_3^2 + 2m_1 \frac{m_2^2 + m_3^2 - M_B^2}{2m_2}\right].\label{eq3}
\end{align}
Here, $M_A$, $M_B$, and $M_C$ represent the masses of the external particles $A$, $B$, and $C$, respectively, as shown in Fig.~\ref{cc1}.
Particle $B$ originates from the decay of the intermediate particle 2 through the reaction $2 \to 3 + B$, implying that the mass of particle 2
must exceed the threshold mass of the 3 and $B$ system.

Based on the mass relationships of particles \( A \) and \( C \), we find that to observe a triangular singularity signal experimentally, the masses of particles \( A \) and \( C \) must meet or exceed the thresholds \( m_1 + m_2 \) and \( m_1 + m_3 \), respectively, as depicted in Fig.~\ref{cc1}. Here, particle \( C \) corresponds to the experimentally identified state \( G(3900) \), which we propose to be a molecular resonance composed of \( \bar{D}D^{*}/D\bar{D}^{*} \), with intermediate particles 1 and 3 representing the molecular constituents of \( G(3900) \). The measured mass centroid of \( G(3900) \) at 3872.5 MeV fulfills the threshold condition specifically for the channels \( \bar{D}^{*0}D^0/D^{*0}\bar{D}^0 \) (\( m_{D^{*0}} = 2006.85 \) MeV, \( m_{D^0} = 1864.84 \) MeV). Given the larger experimental error associated with the mass of \( G(3900) \), its molecular constituents could also be considered as \( D^{*+}D^{-}/D^{*-}D^{+} \) (\( m_{D^{*\pm}} = 2010.26 \) MeV, \( m_{D^{\pm}} = 1869.66 \) MeV). Under these conditions, the mass of \( G(3900) \) must surpass 3879.92 MeV. This finding indicates that the triangular singularity mechanism not only facilitates the identification of \( G(3900) \) as a true \( D^{*}\bar{D}/\bar{D}^{*}D \) molecular state but also enhances precision in mass measurements, as elaborated in Ref.~\cite{Guo:2019qcn}. Thus, utilizing the triangular singularity allows us to further refine the mass range of \( G(3900) \).

\begin{figure}[http]
\begin{center}
\includegraphics[bb=-10 220 550 420, clip, scale=0.80]{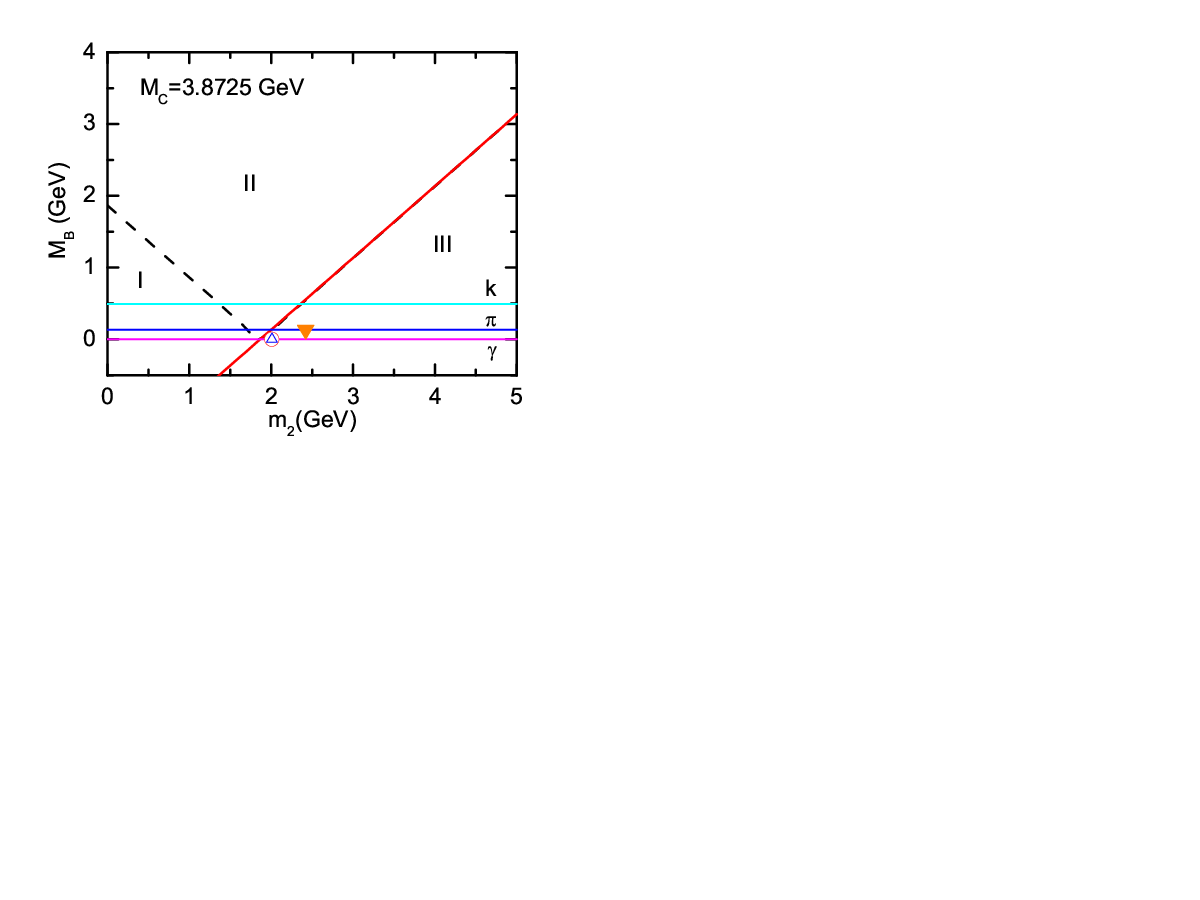}
\caption{The mass ranges of particle 2 and \(B\) satisfy the triangular singularity conditions, which occur only in region III. The red line corresponds to \( M_B = m_2 - m_3 \),
and the black dashed line represents the maximum value from Eq.~\ref{eq3}. Solid lines in magenta, blue, and cyan denote the mass lines for the
$\gamma$, $\pi$, and $K$ meson media, respectively. The red circles, blue triangles, and orange inverted triangles indicate particle 2 as $D^{*0}$,
$D^{*+}$, and $D_1(2410)$, respectively.
}\label{cc2}
\end{center}
\end{figure}

Indeed, as indicated by Eq.~\ref{eq3}, the mass of particle \( C \) achieves a maximum value contingent upon the masses of the intermediate particles and external particle \( B \). This dependency aids in constraining the masses of particle 2 and particle \( B \). For clarity, by substituting the experimentally determined mass of \( G(3900) \) (e.g., \( M_C = 3872.5 \, \text{MeV} \)) into Eq.~\ref{eq3} and designating particles 1 and 3 as \( D^{*0} \) and \( \bar{D}^0 \) respectively, we can derive mass constraints for particles 2 and \( B \), as illustrated in Fig.~\ref{cc2}. The triangular singularity emerges uniquely within region \( III \). From Fig.~\ref{cc2}, it is observable that both the masses of particles 2 and \( B \) attain minimum values, suggesting that the mass range of particle \( C \) is minimized, allowing for an enhanced accuracy in measuring the mass of \( G(3900) \). This also facilitates a more precise determination of the mass of particle \( A \). A plausible explanation is that, at this juncture, the difference between the calculated \( M_B \) and \( m_2 \) is closely aligned with \( m_3 \)~\cite{Liu:2015taa}.

A critical step in advancing this investigation is the identification of an initial particle \( A \) capable of producing a triangular singularity peak that can be directly observed in experiments. The discovery of such a particle would provide strong evidence for the classification of \( G(3900) \) as a legitimate \(\bar{D}D^{*}/D\bar{D}^{*}\) molecular resonance. In the following discussion, we examine several potential initial state particles \( A \) that may engage in the reaction illustrated in Fig.~\ref{cc1}, thereby facilitating the formation of a triangular singularity suitable for experimental validation.

\textit{Results and Discussions}--As illustrated in Fig.~\ref{cc2}, when the mass of the \(G(3900)\) is set to the experimentally determined central value \(M_C = 3872.5 \, \text{MeV}\), and the particle assignments for particles 1 and 3 are \(D^{*0}\) and \(\bar{D}^0\) respectively, we identify that particle \(B\) can be a photon. In this configuration, the minimum mass of particle \(m_2\) is equivalent to \(m_3 = 1864.84 \, \text{MeV}\). A comprehensive search of the Particle Data Group (PDG) reveals that, among the heavy quark particles with mass greater than \(1864.84 \, \text{MeV}\), the only candidate for decaying into the final state \(\gamma D^0\) is \(D^{*0}\), which possesses a substantial branching ratio of \(35.3 \pm 0.9\%\)~\cite{ParticleDataGroup:2022pth}. Substituting these values into Eq.~\ref{eq2} yields an initial state particle mass range of \(M_A = 4013.7 - 4016.4 \, \text{MeV}\), a range capable of generating a triangular singularity. Currently, there remains uncertainty regarding the existence of such a particle that can decay into the final state \(D^{*0} \bar{D}^{*0}\). Notably, in 2022, the Belle Collaboration reported measurements of the cross-section for the two-photon process \(\gamma \gamma \to \gamma \psi(2S)\) across the threshold to 4.2 GeV, identifying a new state with a mass of \(M = (4014.3 \pm 4.0 \pm 1.5) \, \text{MeV}\) and a width of \(\Gamma = 4 \pm 11 \pm 6 \, \text{MeV}\)~\cite{Belle:2021nuv}. However, it is important to note that the global significance of this observation is only \(2.8\sigma\), underscoring the necessity for further experimental data analysis.

It is worth noting that there have been proposals to precisely measure the mass of \( X(3872) \) using triangular singularities in the
\( e^{+}e^{-} \to \gamma X(3872) \) reaction, where a clear peak is expected in the range of \( M_A = 4013.7 - 4016.4 \, \text{MeV} \)
~\cite{Guo:2019qcn,Braaten:2019gfj}.  It is captivating that the latest results from the BESIII Collaboration have confirmed the existence
of \( X(3872) \) in this reaction, providing an opportunity for precise mass determination ~\cite{BESIII:2023hml}. Notably, the BESIII
Collaboration has also reported the discovery of a new particle near the $D^*\bar{D}$ threshold, whose may be the $G(3900)$ state.

We propose to further investigate the reaction $e^+e^- \to \gamma G(3900)$ to disentangle the influence of $X(3872)$ and confirm that the second resonance observed by the BESIII Collaboration corresponds to $G(3900)$. Within the molecular state framework, we identify $G(3900)$ as a $D^{*+}D^{-}/D^{*-}D^{+}$ composite state, with particles 1 and 3 in Fig.~\ref{cc1} corresponding to $D^{*+}$ and $D^{-}$, respectively, and particle 2 identified as $D^{*-/+}$ (Fig.~\ref{cc2}). This interpretation implies that the mass of $G(3900)$ must exceed the central value $M_C = 3872.5$ MeV, yet remain within the calculated range of $M_C = 3879.92 - 3882.47$ MeV, consistent with the experimental uncertainty for $G(3900)$ reported by the BESIII Collaboration. The mass of the initial state $A$, $M_A = 4020.52 - 4023.16$ MeV, allows for the observation of a triangular singularity peak, which is interestingly consistent with the mass of the experimentally observed $X(4020)$, a potential $S$-wave $D^{*}\bar{D}^{*}$ molecule~\cite{BESIII:2013mhi,BESIII:2013ouc,BESIII:2014gnk}. This provides a promising avenue for searching for $G(3900)$.

\begin{figure}[http]
\begin{center}
\includegraphics[bb=-5 225 500 415, clip, scale=0.83]{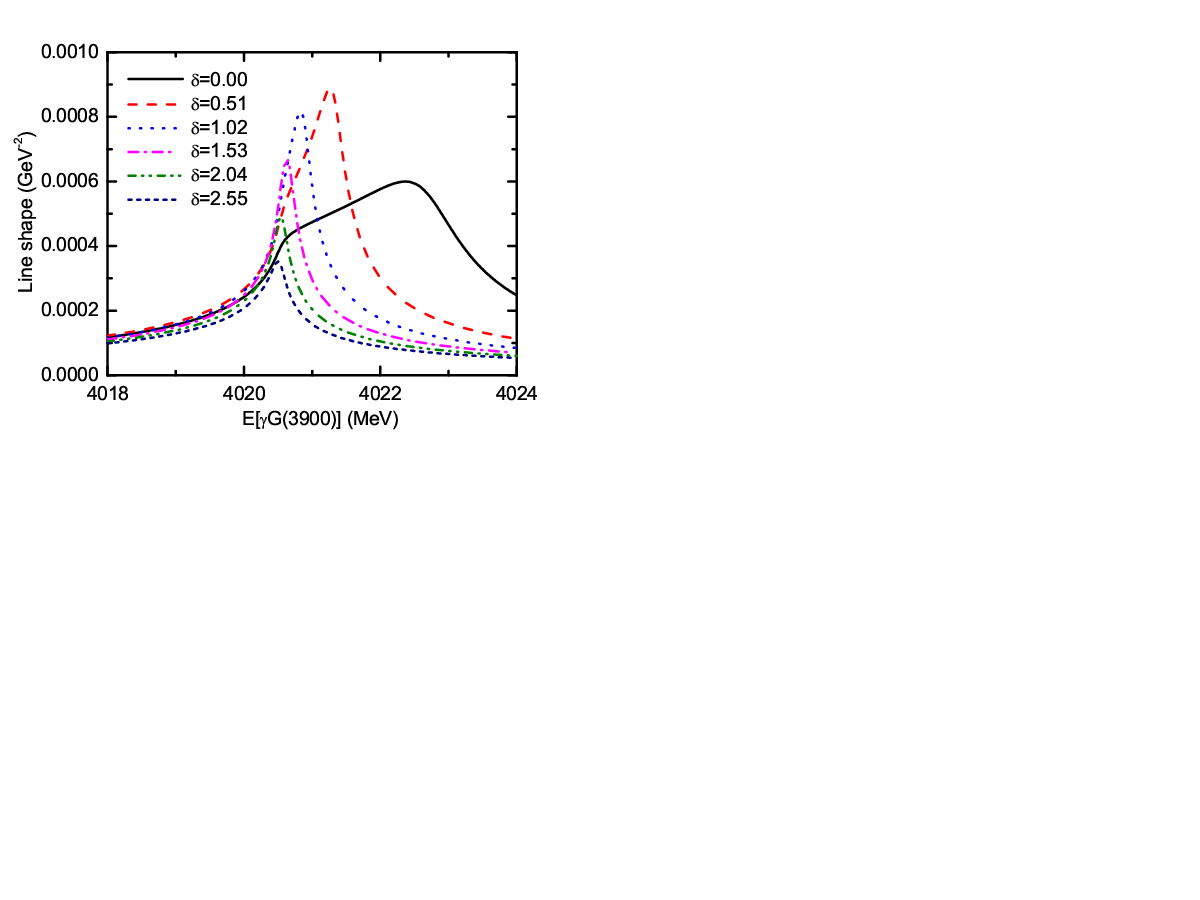}
\caption{The line shapes for the production of \( \gamma G(3900) \) in the \( e^+e^- \to \gamma G(3900) \) reaction through triangular singularity mechanisms.
\(\delta\) represents the energy relative to the \( \bar{D}^{*+}D^{-} \) threshold, but within the range of \( M_C = 3879.92 - 3882.47 \, \text{MeV} \).}\label{cc3}
\end{center}
\end{figure}

The cross section for $e^+e^-$ annihilation producing $X(3872)$ has been carefully calculated in many works, especially in Refs.~\cite{Braaten:2019gwc,Braaten:2019gfj,Dubynskiy:2006cj,Braaten:2020iye,vonDetten:2024eie}, using a triangle diagram that includes the reaction amplitude for $e^+e^- \to D^*\bar{D}^*$
with the production of the $D^*\bar{D}^*$ molecules, which could correspond to the $X(4020)$ and the experimentally unconfirmed $X(4014)$.  Their results show
that the peaks observed in the process mainly arise from the triangle diagram decay of the $D^*\bar{D}^*$ molecule (see Fig.~2 in Ref.~\cite{Braaten:2020iye}),
giving rise to a triangular singularity.  Since we consider that the only difference between $X(3872)$ and $G(3900)$ is that one is an $S$-wave $D\bar{D}^{*}$ molecule and the other is a $P$-wave, we evaluate the linear shape of the $e^+e^- \to \gamma G(3900)$ reaction in this work using the same formalism.  The cross section for $e^+e^- \to \gamma G(3900)$ reaction can be approximately considered as proportional to ${\cal{F}}^2(\bar{D}^{*}D^{*})$, where ${\cal{F}}$ is the amplitude for the $\bar{D}^{*}D^{*}$ molecule $X(4020)$ decaying into the final state $\gamma G(3900)$ through the triangle diagram (see Fig.~\ref{cc1}), which can give rise to a sharp peak.  In the non-relativistic limit, ${\cal{F}}$ has been computed in several work~\cite{Guo:2019qcn,Braaten:2019gwc,Braaten:2019gfj,Braaten:2020iye,vonDetten:2024eie}, and we adopt a simple form from Ref. ~\cite{Guo:2019qcn},
\begin{align}
{\cal{F}}\propto\frac{1}{E_{\gamma}}[\arctan{(\frac{c_2-c_1}{2b\sqrt{c_1}})}+\arctan{(\frac{c_1-c_2+2b^2}{2b\sqrt{c_2-b^2}})}]\label{eq4},
\end{align}
where $E_{\gamma} = (E_{G\gamma}^2 - m_G^2)/(2E_{G\gamma})$ is the photon energy, $E_{G\gamma}$ is the invariant mass of the $G(3900)\gamma$ system,
$b = m_{D^{*+}}E_{\gamma}/(m_{D^{-}} + m_{D^{*+}})$, $c_1 = m_{D^{*+}} \left( 2m_{D^{*+}} - E_{X\gamma} - i\Gamma_{D^{*+}} \right)$, and
$c_2 = \mu_{D^{*+}} \left[ 2 \left( m_{D^{*+}} + m_{D^{-}} + E_{\gamma} - E_{X\gamma} \right) + E_{\gamma}^2/m_{D^{-}} - i\Gamma_{D^{*+}} \right]$.
Here, $m_G$, $m_{D^{*+}}$ and $m_{D^{-}}$ are the masses of the $G(3900)$, $D^{*+}$ and $D^{-}$, respectively, $\mu_{D^{*+}}  = m_{D^{*+}} m_{D^{-}}/(m_{D^{*+}} + m_{D^-})$
is the reduced mass of $D^{*+}$ and $D^{-}$, and $\Gamma_{D^{*+}}=83.4$ KeV is the $D^{*+}$ width.  Note that Eq.~\ref{eq4} neglects the spin structures of the particles, as they do not affect the triangular singularity peak~\cite{Liu:2015taa,Guo:2010ak}, and the use of \(\propto\) instead of the equals sign indicates the omission of the coupling constants.
  We plot \( \sigma[e^+e^- \to \gamma G(3900)] \simeq {\cal{F}}^2 \) in Fig.~\ref{cc3}, where prominent peaks are identified, providing direct evidence for \( G(3900) \) as a \( \bar{D}^{*}D \) molecule, which, once detected experimentally, would confirm its existence.

Replacing the photon with a \(\pi\), then according to Eqs.~\ref{eq2} and \ref{eq3}, only more precise measurements of \( G(3900) \)
and the mass of particle \( A \) (identified as either \( X(4014) \) or \( X(4020) \)) are needed. As such, further discussion on this case is unnecessary. However, by interchanging the intermediate particles 1 and 3, assigned as $\bar{D}^0$ and $D^{*0}$, we find that a heavier quark state with a minimum mass of 2141.83 MeV is required to decay into the $\pi D^{*}$ channel. Notably, despite the presence of numerous particles that can decay into $\pi D^{*}$, the decay branching ratio of $D_1(2420)$ into $\pi D^{*}$ is well-established and dominant in this channel, supported by fits to experimental data in Ref.~\cite{BESIII:2023hml} and theoretical analyses in Ref.~\cite{Colangelo:2005gb}. This substantial $D_1(2420) \to \pi D^{*}$ branching ratio, in turn, contributes significantly to a substantial production cross-section for $G(3900)$, as noted in Ref.~\cite{Braaten:2019gfj}.

We extend our earlier analysis to determine the initial state mass required to generate a triangular singularity peak, which is found to lie within the range $M = 4286.94 - 4303.62$ MeV~\cite{Huang:2024asn}. Of particular interest is the observation of the $Y(4320)$ resonance by the BESIII collaboration in the $e^{+}e^{-} \to \pi^{+}\pi^{-} J/\psi$ channel, with measured properties of $M = 4298 \pm 12 \pm 26$ MeV and $\Gamma = 127 \pm 17 \pm 10$ MeV~\cite{BESIII:2022qal}. Given its experimental observation in the $\pi^{+}\pi^{-} J/\psi$ final state, it is plausible that this resonance contains a $c \bar{c} q \bar{q}$ component and can decay into $D_1 \bar{D}$.  Notably, previous research suggests that this state may correspond to a higher partial-wave molecular structure of $D_1 \bar{D}$~\cite{Huang:2024asn}, primarily due to the experimentally observed $Y(4230)$, which is widely regarded as an $S$-wave $D_1 \bar{D}$ molecular state~\cite{Ji:2022blw, Wang:2013cya}. We propose experiments to probe $G(3900)$ near the $D_1 \bar{D}$ threshold through the $e^+e^- \to \gamma G(3900)$ reaction, which will result in the production of a triangular singularity peak with a mass range between $M_A = 4286.94 - 4303.62$ MeV.  Since the linear shape is similar to Fig.~\ref{cc3}, we did not plot it.

\textit{Summary}--The recently identified exotic hadron, $G(3900)$~\cite{BESIII:2024ths}, is situated near the mass of the well-known $X(3872)$. Theoretical investigations suggest that $G(3900)$ may represent a high orbital angular momentum molecular state characterized by $\bar{D}D^*/D\bar{D}^*$ components~\cite{Lin:2024qcq}. However, the precise nature of this state continues to be a subject of debate, with some analyses proposing that the $G(3900)$ might not be a true hadronic state, but rather an artifact arising from the underlying reaction dynamics~\cite{Eichten:1979ms, Salnikov:2024wah, Husken:2024hmi}. To clarify this issue, comprehensive experimental studies across multiple reaction channels are essential to ascertain whether the $G(3900)$ constitutes a genuine hadronic state and, if so, whether it is associated with a $P$-wave $\bar{D}D^*/D\bar{D}^*$ molecular resonance.

In this Letter, we propose leveraging the triangular singularity mechanism to investigate the $G(3900)$ in reaction channels beyond the $e^+ e^- \to \bar{D}D$ process. The triangular singularity mechanism is capable of producing distinctive resonance peaks, thus serving as a powerful experimental method to evaluate the nature of the $G(3900)$ and to further explore its molecular state characteristics. Our analysis indicates that, if the $G(3900)$ is treated as a legitimate state interpretable as a $\bar{D}D^*/D\bar{D}^*$ molecular resonance, the only relevant states capable of generating triangular singularity peaks through triangle diagram decays to the final states $\gamma G(3900)$ or $\pi G(3900)$ are $X(4020)$, $Y(4320)$, and the not-yet-fully-confirmed $X(4014)$. These states can be recognized as $\bar{D}^*D^*$, $\bar{D}^*D^*$, and $\bar{D}_1D$ molecules, respectively, and can be produced in $e^+ e^-$ collisions. Consequently, we recommend experiments focusing on the $e^+ e^- \to \gamma (\pi) G(3900)$ reactions near the $\bar{D}^*D^*$ and $\bar{D}_1D$ thresholds to ascertain whether the $G(3900)$ is indeed a genuine $\bar{D}^*D/D^{*}\bar{D}$ molecular state.

\section*{Acknowledgments}

This work is supported by National Key R\&{}D Program of China No.2024YFE0109800 and 2024YFE0109802.

\end{document}